\documentclass[sigconf,nonacm]{acmart}

\usepackage{amsmath}

\usepackage{amssymb}
\usepackage{threeparttable}

\graphicspath{{figures/}}
\definecolor{steelblue}{RGB}{70,130,180}
\setcopyright{none}
\hypersetup{
  hidelinks,
  pdfauthor={Jingqiu Ding and Daojing Zhai},
  pdfsubject={},
  pdfcreator={},
  pdfproducer={},
  pdftitle={Beyond Anonymous Order Flow: Wallet Identity as a Feature for Short-Horizon Return Prediction}
}

\title[Beyond Anonymous Order Flow]{Beyond Anonymous Order Flow: Wallet Identity as a Feature for Short-Horizon Return Prediction}

\author{Jingqiu Ding}
\orcid{0009-0001-2142-0501}
\affiliation{%
  \institution{ETH Zurich}
  \city{Zurich}
  \country{Switzerland}}
\email{jingqiu.ding@inf.ethz.ch}

\author{Daojing Zhai}
\orcid{0000-0002-0717-9707}
\affiliation{%
  \institution{Baruch College, City University of New York}
  \city{New York}
  \country{USA}}
\email{daojing.zhai@baruch.cuny.edu}
\renewcommand{\shortauthors}{Ding and Zhai}

\begin{document}

\begin{abstract}
Conventional public order-book data summarize trading without identifying the accounts behind it.
Hyperliquid's Level-4 record also exposes persistent pseudonymous wallet addresses. We reconstruct
full-depth books from consensus-ordered messages and evaluate wallet identity as a feature for
short-horizon return prediction. The July 1--27, 2026 record covers 17.1~billion messages,
14.3~million aggressive orders by 147,113 wallets, and \$84.3~billion in taker notional across ten
markets. In the three model markets, notional-weighted ten-second markout rankings persist across
adjacent ten-day windows (Spearman $\rho=0.52$); covariate-adjusted validation markouts reach
3.11 basis points in the top ventile. On a 100-millisecond grid, eleven features describing
previously scored top-decile wallets raise one-second out-of-sample $R^2$ from 10.88\% to 12.31\%
under ridge, a 13.2\% relative increase ($t=9.2$). The increment exceeds those of 200
activity-matched cohorts through ten seconds, remains positive under 200--300\,ms feature delays,
and is positive in an earlier December 2025 sample under the same volatility-scaled specification.
This retrospective evaluation supports incremental prediction relative to the tested anonymous
feature set. Specification selection, carried-forward states during data gaps, and consensus-time
rather than receipt-time indexing limit confirmatory and deployment interpretations.

\end{abstract}

\ccsdesc[500]{Applied computing~Economics}
\ccsdesc[500]{Computing methodologies~Supervised learning by regression}

\keywords{limit order books, market microstructure, adverse selection, trader identity,
short-horizon return prediction, decentralized exchanges}

\maketitle
\hypersetup{
  pdfauthor={Jingqiu Ding and Daojing Zhai},
  pdftitle={Beyond Anonymous Order Flow: Wallet Identity as a Feature for Short-Horizon Return Prediction},
  pdfkeywords={limit order books, market microstructure, adverse selection, trader identity, short-horizon return prediction, decentralized exchanges}
}

\section{Introduction}
\label{sec:introduction}

Short-horizon return prediction from limit-order-book data is a supervised learning problem
with established predictors: signed order flow, depth imbalance, recent returns, realized
volatility, and the quoted spread \citep{cont2014price,sirignano2019universal,kolm2023deep}.
These aggregate variables summarize what the market is doing. Persistent account identifiers
also allow a model to condition on whose activity it observes. Adverse-selection models motivate
this distinction: counterparties differ in the price movements following their trades
\citep{bagehot1971only,kyle1985continuous,glosten1985bid}. Conventional public exchange feeds
usually do not expose account histories, whereas wallet-attributed decentralized feeds can.

Hyperliquid runs a central limit order book with price--time priority and publishes
consensus-ordered order submissions, cancellations, rejections, and fills with wallet addresses
\citep{albers2026openbook}. We use this record to ask two questions. Do wallet markout rankings
persist across disjoint windows? Does the activity of previously high-ranked wallets improve
short-horizon return forecasts beyond a literature-based anonymous benchmark?

\paragraph{Pipeline and evaluation.} A non-validating node records committed events. A deterministic
replay rebuilds the book, pairs maker and taker fills, and links orders to wallets. Ten-second
oracle observations gate emission of reconstructed state when large midpoint discrepancies occur;
the gate does not certify every resting order. The July 1--27 record contains 17.1~billion
messages, 14.3~million aggressive orders, and \$84.3~billion in taker notional across ten
perpetual markets (Table~\ref{tab:data-summary}). Estimation uses \textsc{btc}, \textsc{eth},
and \textsc{sol}.

We score wallets using July 1--10 aggressive-order markouts, fit forecasting models on July 11--20,
and evaluate them on July 21--27. A ten-variable anonymous block summarizes prices, quotes, and
flow; eleven additional variables summarize previously scored wallets' activity. Ridge and
gradient-boosted trees use the same feature definitions. Wallet scores and model parameters use
earlier observations than their evaluation targets. The final cohort cut, volatility scaling,
and reported evaluation window were selected retrospectively, however, so the study is a
chronological out-of-sample comparison rather than a prospectively locked test.

\paragraph{Findings.} Wallet scores have an across-window rank correlation of 0.52, and adjusted
validation markouts rise sharply in the upper tail, reaching 3.11 basis points in the top ventile.
At one second, identity raises ridge $R^2$ from 10.88\% to 12.31\% ($+13.2\%$ relative,
$t=9.2$). Trees extract more from the anonymous block (19.48\%) and still gain from identity,
reaching 20.65\% ($+6.0\%$, $t=5.0$). Relative increments are largest at short horizons and
remain positive under the top quintile. A peer-adjusted score, controlling for
same-market, same-minute markouts, yields a smaller 6.8\% one-second increment.

\paragraph{Controls and scope.} The one-second ridge increment is 1.6 times the largest increment
among 200 cohorts matched on scoring-window notional and order count. Donor shortages make these
cohorts smaller than the toxic cohort, so their ranking is descriptive. Delaying every feature by
200 and 300 milliseconds leaves relative ridge gains of 10.2\% and 9.1\%, respectively. An
earlier public December 2025 sample also gives a positive increment under the same volatility-scaled
specification. These exercises test selected activity controls and simulated delays; they do not
identify private information, eliminate all anonymous representations, or demonstrate executable
profit.

\paragraph{Contributions.} We reconstruct full-depth books with wallet attribution and coverage checks, measure the persistence of aggressive-order markout rankings, and quantify the
forecasting increment from wallet-specific quotes and flow. Drop-one refits locate much of the
increment in cohort quote updates and distinct-buyer breadth. The empirical contribution is the
evaluation of persistent wallet histories, building on the existing Level-4 data literature.

\paragraph{Related work.} Because the canonical models leave the counterparty unidentified,
adverse selection has been inferred from anonymous data: spread decompositions
\citep{glosten1988estimating}, permanent price impact \citep{hasbrouck1991measuring}, and the
structurally estimated probability of informed trading \citep{easley1996liquidity}. These measures
can fail exactly when identification matters most \citep{collindufresne2015do}. Where identifiers
have been observable they matter: anonymity changes liquidity
\citep{foucault2007does,meling2021anonymous}, visible identifiers are learned from
\citep{linnainmaa2012lack}, account-level records show performance and information to be persistent
trader attributes \citep{kirilenko2017flash,baron2019risk}, and a leaked client identity is worth
trading against \citep{barbon2019brokers}. These traditional account-level studies generally rely on identifiers visible to a
broker, regulator, or researcher rather than to all users of a public feed. Crypto microstructure has studied arbitrage and
market quality \citep{makarov2020trading}, the strategic games public blockchains enable
\citep{daian2020flash}, perpetual futures \citep{he2024fundamentals}, and wallet-tagged insider
activity \citep{felezvinas2022insider}; on Hyperliquid's Level-4 record specifically,
\citet{albers2026openbook} introduce the data, \citet{albers2026priceimpact} show that rejected
orders predict returns, and \citet{barone2026sunshine} that disclosed execution intentions lower
trading costs, while \citet{zhang2025clusterlob} recover trader types by clustering NASDAQ order
events. Optimal quoting conditions on inventory and adverse selection \citep{avellaneda2008high}
and, inside dealers, on client identity \citep{cartea2026detecting}, whereas
anonymous forecasting models cannot construct wallet histories at all
\citep{cont2014price,sirignano2019universal,kolm2023deep}. We shift the unit of analysis to the
persistent wallet and measure its incremental forecasting content directly.

\section{Setting and Data Pipeline}
\label{sec:data}

Hyperliquid is a decentralized exchange for perpetual-futures trading. It runs on
HyperCore, which maintains a central limit order book per contract and
matches on price--time priority. Order submissions and cancellations are sent to the network's
validators, collected into short consensus-ordered \emph{blocks}, and each block is then executed
against the relevant book. Once a block commits, its trades, positions, and resulting book state
are final, so the block is the atomic unit of ordering and of information updates.

\begin{figure*}[tp]
    \centering
    \includegraphics[width=0.9\textwidth]{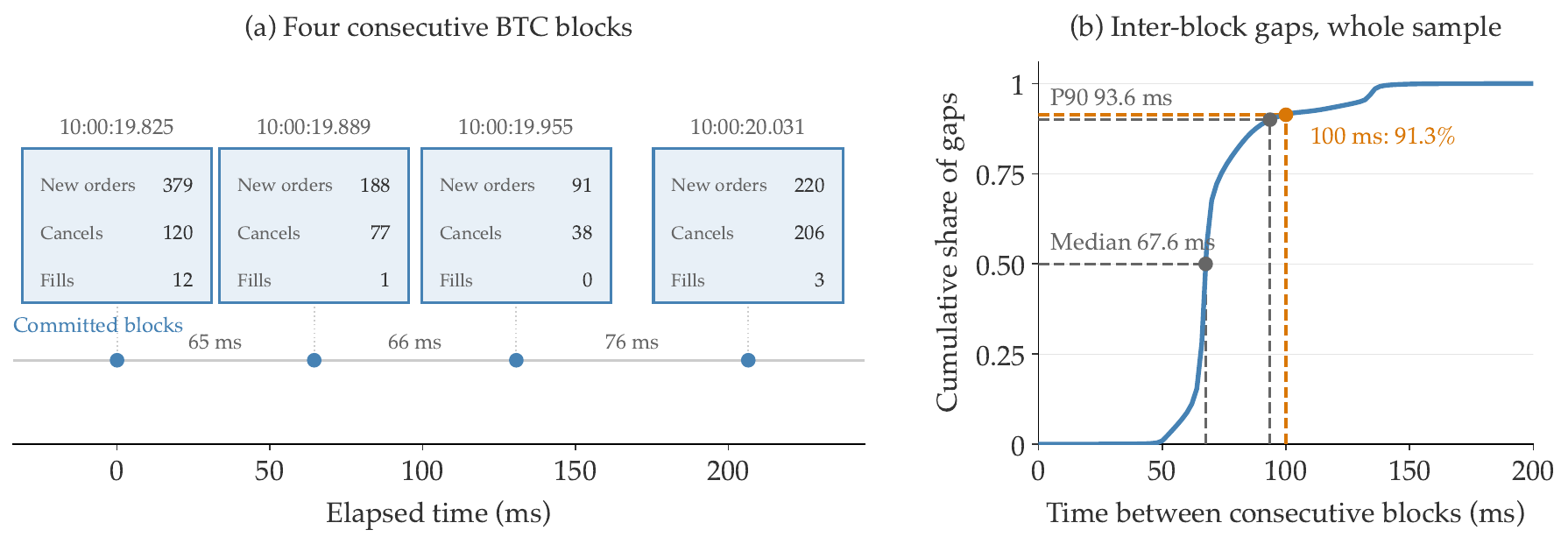}
    \caption{Committed blocks and inter-block timing. (a) Four consecutive \textsc{btc} blocks
    on July 15, 2026, at their consensus timestamps in New York time, with the events
    each carries. (b) ECDF of all
    22{,}734{,}866 inter-block gaps in the sample. Block timestamps are truncated to the
    millisecond; gaps are computed at full precision.}
    \Description{Panel (a) shows four block boxes with counts of new orders, cancellations and
    fills, spaced 65, 66 and 76 milliseconds apart. Panel (b) shows the cumulative distribution of
    inter-block gaps, with a median of 67.6 milliseconds and 91.3 percent of gaps at or below 100
    milliseconds.}
    \label{fig:blocks}
\end{figure*}

Blocks arrive at irregular calendar-time intervals, motivating the sampling design downstream. Figure~\ref{fig:blocks} makes that clock
concrete: Panel~(a) places four consecutive \textsc{btc} blocks at their consensus timestamps with
the events each carries, and Panel~(b) gives the distribution of all 22{,}734{,}866 inter-block
gaps. The median gap is 67.6~milliseconds, the 90th percentile is 93.6, and 91.3\% of gaps do not
exceed 100~milliseconds. At any instant the current midpoint is the close of the most recent block
committed by that time; this committed-block clock underlies every price and return in the paper.

The distinguishing feature of the venue is that all of this is public \citep{albers2026openbook}.
Every committed block records every order attempt, whether a submission, a cancellation, or a
rejection, together with every book update and execution. We collect these records continuously
through a non-validating node on a dedicated cloud server, which connects to the peer-to-peer
network through bootstrap peers, synchronizes the committed chain, and writes unbuffered output
locally. In parallel we query the \texttt{metaAndAssetCtxs} endpoint every ten seconds for mark
prices, oracle prices, and open interest; the oracle ticks provide an external check on the
reconstruction. Replaying the messages in consensus order produces a Level-4 database of
order-by-order book states, reconstructed order lifecycles, rejected attempts, matched maker and
taker executions, and the persistent pseudonymous wallet address attached to each order and
trade.\footnote{We use ``identity'' throughout for this address-level behavioral identifier. One
entity may control several wallets, and one wallet may aggregate several principals; the analysis
does not identify legal persons.}

Reconstructed state is emitted only while it passes a gate against the contemporaneous oracle, which
requires twelve consecutive ten-second ticks to agree with the reconstructed midpoint to within 25
basis points. The gate suppresses emission while these midpoint checks fail. The gate is
closed for 0.97\% of coin-time over July 1--27 and 1.83\% inside the evaluation window, and 81\% of
suppressed time is common to all three markets at once, consistent with a shared oracle-side outage, although not sufficient to identify its cause.\footnote{Three further checks. \emph{Fill
pairing:} each of the 27{,}902{,}380 crossed fills on July 1--27 pairs with a maker-side record at
the same (day, market, trade identifier), with zero unmatched legs by count and by notional.
\emph{Segment seams:} media damage and node restarts split the replay into thirteen segments;
replaying July 12 standalone, from an empty book through warm-up to gate re-admission, reproduces the continuous
replay's bid, ask, mid, and spread on 99.994\% of 1.32 million comparable blocks. \emph{Coverage:}
twenty-two of the thirty days in the July 1--30 collection window, which is wider than the
sample, show a maximum intra-day block gap of three or fewer; every
fitting-window day satisfies that bound; the evaluation window does apart from a sixteen-block gap
on July 25 and a 28{,}326-block gap on July 27; and the scoring window does apart from July 1 and
July 2, which carry gaps of 27{,}900 and 59{,}415 blocks, and July 3, 4 and 6, whose maxima are
four, seven and four blocks.} This is a coarse midpoint check: the 25\,bp tolerance is larger than the markout effects
studied here, and resting-order errors can survive while the midpoint agrees.

All estimation uses \textsc{btc}, \textsc{eth}, and \textsc{sol}, the three markets with the
greatest message activity; Table~\ref{tab:data-summary} reports them against the aggregate over all
ten most active perpetuals, which \textsc{btc} leads with \$46.8~billion of the \$84.3~billion in
taker notional.

\begin{table}[tbp]\centering
\begin{threeparttable}
\caption{Level-4 sample, July 1--27, 2026}
\label{tab:data-summary}
\fontsize{7.7}{8.8}\selectfont\setlength{\tabcolsep}{3pt}\renewcommand{\arraystretch}{1.05}
\begin{tabular}{@{}lrrrrrrr@{}}
\toprule
Market & Msgs & BBO & Orders & Fills & Wallets & Notional & Spread \\
 & (bn) & (m) & (m) & (m) & (000s) & (\$bn) & (bps) \\
\midrule
BTC & 9.75 & 0.94 & 4.02 & 8.66 & 98.9 & 46.8 & 0.20 \\
ETH & 3.03 & 0.91 & 2.04 & 3.89 & 61.7 & 18.7 & 0.62 \\
SOL & 1.35 & 1.69 & 1.35 & 2.16 & 50.5 & 4.9 & 0.26 \\
\midrule
All ten & 17.09 & 27.26 & 14.27 & 27.90 & 147.1 & 84.3 & 0.92 \\
\bottomrule
\end{tabular}
\begin{tablenotes}[flushleft]\fontsize{6.6}{7.7}\selectfont
\item \textit{Notes:} The three model markets used for all estimation, above the aggregate over the
ten most active Hyperliquid perpetuals by L4 message count; the other seven enter that aggregate
but are not listed. L4 messages combine order-status and book-diff records; BBO updates are
gate-valid blocks in which the best quote price changed; taker fills are individual legs, grouped into
one aggressive order when they share a wallet, order identifier, and market. Spread is the mean
quoted spread immediately before aggressive orders. Wallets are deduplicated across markets in the
All row.
\end{tablenotes}
\end{threeparttable}\end{table}

\section{Scoring Wallets by Markout}
\label{sec:toxicity}

Every action on Hyperliquid carries the pseudonymous address of the account behind it, so a
wallet's history can be linked to its subsequent activity. We build the score from aggressive
orders, because there the trader chooses immediate execution and crosses the spread, and canonical
adverse-selection models read aggressive flow as revealing beliefs about future prices
\citep{bagehot1971only,kyle1985continuous,glosten1985bid}. Motives differ: a directional trader
expresses a view, while a market maker hedges inventory or responds to imbalance. In either case the
direction of the resulting trade may carry information about subsequent prices. The working
hypothesis is therefore that wallets can be ranked by the signed markouts of their past aggressive
trades, and that subsequent activity by high-ranked wallets helps predict returns.

The design needs persistence along two dimensions. Addresses must remain observable long enough to
accumulate a history: of the 2{,}314 wallets with at least 100 qualifying aggressive orders during
July 1--10, 86.9\% trade again during July 11--20, and those continuing wallets account for 50.3\%
of all discretionary (non-TWAP, non-liquidation) taker notional in the first window and 43.6\%
in the second. Informativeness
itself must also persist, which Section~\ref{sec:toxic-results} tests.

\subsection{Definition}
\label{sec:measurement}

Following the microstructure literature, we measure post-trade adverse selection with the signed
midpoint markout \citep{barone2026sunshine,cartea2026detecting}. For an aggressive order $e$ filled
at time $t_e$ with notional $n_e$ and direction $q_e$ ($+1$ buy, $-1$ sell), the ten-second
markout, in basis points of the pre-event midpoint, is
\begin{equation}
  x_e = 10^4\, q_e\,\frac{m_{t_e+10s}-m_{t_e^-}}{m_{t_e^-}},
  \label{eq:advselection}
\end{equation}
where $m_{t_e^-}$ is the midpoint at the close of the last block strictly before the event's block,
and $m_{t_e+10s}$ is the latest midpoint committed by $t_e+10s$. A positive markout means the later midpoint lies in the aggressor's direction relative to
the pre-block midpoint. The measure includes contemporaneous block impact as well as later moves. ``Toxicity'' is a descriptive label for heterogeneity in this quantity; it is not a
claim that a wallet is one trader or trades on private information.

The wallet score is the notional-weighted mean markout over a frozen ten-day window,
\begin{equation}
  \alpha_w=\frac{\sum_{e\in w} n_e\, x_e}{\sum_{e\in w} n_e},
  \label{eq:wallet-toxicity}
\end{equation}
computed over July 1--10 for wallets with at least 100 qualifying aggressive orders, excluding TWAP
and liquidation events. Using an earlier scoring window separates the score from July evaluation targets. Among the 2{,}314 scored wallets, the top decile (231 wallets) is the
primary toxic cohort and the top quintile (462 wallets) an alternative tail cut.

A frozen window has one limitation: it goes stale and never scores wallets that first appear after
July 10. We evaluate a peer-adjusted variant that nets each order's markout against other wallets' notional-weighted mean
markout in the same coin and minute. The adjustment controls for coarse market-by-minute variation and gives a 6.8\% one-second
forecasting gain, roughly half the raw score's 13.2\%. It does not remove common timing
within those minutes. For comparability, raw and peer-adjusted forecasting scores use the same
order universe, requiring another wallet's order in the minute.

\subsection{Does the score persist?}
\label{sec:toxic-results}

\begin{figure*}[tp]
    \centering
    \includegraphics[width=0.9\textwidth]{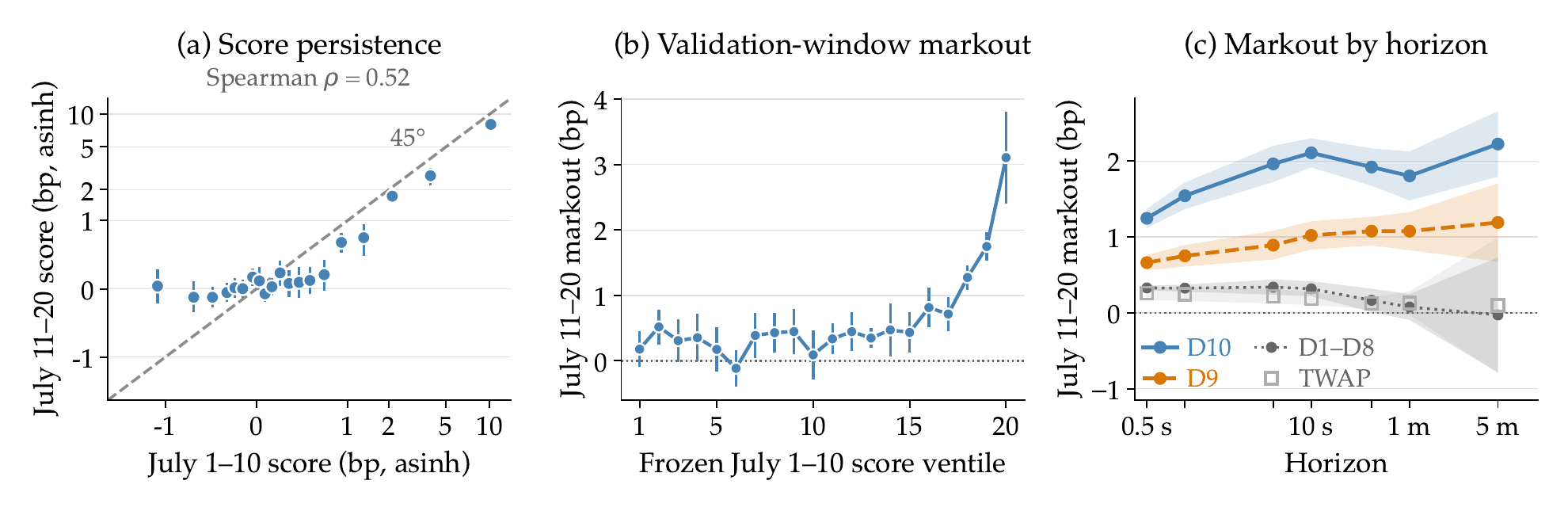}
    \caption{Persistence and horizon profile of wallet toxicity. (a) Scoring- against
    validation-window score. (b) Adjusted validation markout by scoring-window ventile.
    (c) Markout by horizon for D10, D9, D1--D8, and native TWAP child orders. Bars and bands show
    approximate 95\% intervals computed as $\pm 1.96$ standard errors for wallet-level means
    in (a) and means of daily estimates in (b)--(c).}
    \Description{Panel (a) plots July 11 to 20 wallet scores against July 1 to 10 scores by
    ventile, aligned along the 45-degree line with Spearman rho of 0.52. Panel (b) shows a flat
    markout profile through the fifteenth ventile that rises sharply to about 3.1 basis points at
    the twentieth. Panel (c) shows top-decile markouts rising from about 1.25 basis points at half
    a second to above 2 at ten seconds and staying elevated at five minutes; D9 tracks between
    them, rising from about 0.65 to 1.2 basis points, while the D1 to D8 remainder and TWAP child
    orders stay near zero.}
    \label{fig:persistence}
\end{figure*}

We score on July 1--10 and validate on July 11--20, asking whether the scoring-window ranking
still orders validation-window markouts. Among 965 wallets clearing the 100-order threshold in both windows, the rank correlation
is $\rho=0.52$, and
Figure~\ref{fig:persistence}(a) shows the persistence is asymmetric: the bottom decile regresses
toward zero while the top decile stays elevated and close to the 45-degree line. Splitting each window into odd and even calendar days and rescoring the same wallets gives
split-half rank correlations of 0.52 and 0.48. Applying the Spearman--Brown formula
$2r/(1+r)$ gives approximate full-window reliability benchmarks of 0.68 and 0.65. The
across-window correlation is about 78\% of their geometric mean. This is a heuristic rather
than an exact ceiling for rank correlations; interpreting it as measurement reliability requires
parallel measurements and independent errors, which serially dependent trading may violate.

The ventile sort is the direct test of whether high scoring-window ranks predict larger
validation-window markouts. After controlling for market, time, order size, volatility, and spread,
markouts stay flat through the fifteenth ventile before rising sharply to 3.11 basis points in the
top ventile (Panel~(b)). Daily notional-weighted regressions include ventile indicators,
centered log order notional, volatility, spread, market indicators, and eight three-hour buckets;
plotted coefficients average the ten daily estimates. The economic content is concentrated in the upper tail, which is why the
prediction experiment tracks the top decile.

A persistent ranking could still reflect temporary price pressure from the aggressive trades
themselves, in which case the markout should reverse quickly. It does not: the top-decile markout
rises from 1.25 basis points at half a second to 2.11 at ten seconds and remains elevated at five
minutes (Panel~(c)), while the D1--D8 remainder stays near zero. Native TWAP child orders, whose
timing is scheduled rather than chosen per execution, provide a mechanical-execution benchmark; the
D1--D8 remainder tracks it at every horizon whereas the top-decile markout is several times larger
and rises as the TWAP markout decays. This profile is inconsistent with purely transient price
pressure, although persistent impact, repeated same-direction flow, and latent metaorders remain
possible.

Persistence is useful only if the top-ranked wallets keep trading, and they do: 91.3\% of the 231
top-decile wallets trade again during the validation window, the highest continuation rate of any
decile, and that decile alone carries 31.0\% of scoring-window and 25.1\% of validation-window
discretionary taker notional. Across deciles fixed once on the frozen score, the validation markout
stays at or below 0.60 basis points through D8, reaches 1.09 in D9, and doubles again to 2.20 in
D10. The result reflects persistence among continuing wallets rather than attrition among the
initially ranked ones.

\paragraph{Cross-venue latency.} Persistent markouts could arise mechanically if some wallets react
to price changes on other venues before Hyperliquid adjusts, the cross-venue analogue of toxic
arbitrage \citep{foucault2017toxic}. Such a wallet should trade toward the gap between the venue's
external oracle and its local midpoint. The data show the opposite: top-decile orders arrive on the
gap's side only 41.3\% of the time against 57.9\% for the bottom decile, and the mean signed gap
they trade into is negative. The markout also survives two controls. The Fama--MacBeth daily
D10$-$D1 difference in markout moves only from 1.72 to 1.75 basis points once oracle-gap controls
enter the same specification, and the subsample with a gap of at most one basis point still yields
2.12 basis points for the top decile against 0.16 for the bottom. These controls use an oracle
polled every ten seconds and therefore do not rule out sub-second cross-venue arbitrage.

\section{Does Identity Predict Returns?}
\label{sec:prediction}

Section~\ref{sec:toxic-results} establishes that markouts persist across identifiable wallets. This
section asks whether identifying the wallets behind current market activity improves short-horizon
return forecasts beyond what an anonymous order book supplies. The design is a controlled
comparison: an anonymous benchmark and an identity-augmented model observe the same public tape.
The benchmark summarizes aggregate prices, quotes, and trade flow; the augmented model adds
parallel quote and flow measures for top-decile wallets, together with wallet-specific measures of
trading breadth and quoted depth. The difference measures predictive improvement relative to this anonymous representation.

\subsection{Feature blocks}
\label{sec:prediction-features}

The anonymous block combines published predictors from
several strands of the microstructure literature; Table~\ref{tab:part2-variables} maps each
variable to its source. Quote variables capture depth imbalance at the best quotes and within five
basis points of the midpoint, together with quote-update order-flow imbalance over the previous one
and thirty seconds \citep{stoikov2018micro,xu2019multilevel,cont2014price}. We multiply the near-touch pair by the coin's previous-day realized one-second return
volatility, measured in basis points, to include an imbalance--volatility interaction. Flow variables capture
the latest signed print and signed taker notional over the same windows
\citep{hasbrouck1991measuring,biais1995empirical}. Price variables capture the previous one-minute
return, one-minute realized volatility, and the quoted spread
\citep{andersen2003modeling,glosten1988estimating}.

The identity block constructs analogous quote and flow measures from top-decile wallets and
adds four address-based variables: buyer-minus-seller breadth over one and thirty seconds,
the decile's share of best-quote depth, and its contribution to best-quote imbalance. Breadth
is a rolling sum of distinct buyers minus sellers within each 100\,ms bucket; a wallet may
count again in later buckets. Cohort quote-update OFI sums changes in bid-minus-ask USD depth,
whereas anonymous OFI uses the price-indicator transitions of \citet{cont2014price}. Price variables have no identity
counterpart because prices are common to all traders. The reported comparisons use these fixed feature definitions.\footnote{This is a literature-based, non-maximal anonymous benchmark. Richer multi-level
depth, cancellation, rejection, intensity, and cross-venue predictors may reduce the measured
increment; \citet{albers2026priceimpact} show that rejected orders predict returns on this venue.}

Let $B,A$ denote total bid and ask USD depth at the best quotes, and $B_H,A_H$ the cohort's
corresponding depths. Cohort imbalance is $(B_H-A_H)/(B_H+A_H)$, set to zero when the
denominator vanishes. Share and contribution are $(B_H+A_H)/D$ and $(B_H-A_H)/D$, with
$D=\max(B+A,B_H+A_H)$; zero denominators give zero. This clamp bounds both measures despite
transient denominator inconsistencies. Near-touch imbalance uses depth within 5\,bp of the
midpoint. Flow is the sum of signed aggressive-order notional over the stated window; print
is the latest signed order notional.

\begin{table}[tbp]\centering
\begin{threeparttable}
\caption{Predictor sets: the anonymous--identity mapping}
\label{tab:part2-variables}
\fontsize{7.0}{8.1}\selectfont\setlength{\tabcolsep}{2.5pt}\renewcommand{\arraystretch}{1.05}
\begin{tabular}{@{}lllll@{}}
\toprule
Family & Anonymous & Identity & Shared definition & Ref. \\
\midrule
\multicolumn{5}{@{}l}{\textit{Panel A: paired variables (analogous cohort measures)}} \\
\midrule
Quote & \texttt{imb}     & \texttt{hotImb}     & best-quote depth imbalance & \cite{stoikov2018micro} \\
Quote & \texttt{nearImb} & \texttt{hotNear}    & near-touch imbalance\tnote{c} & \cite{xu2019multilevel} \\
Quote & \texttt{ofi1s}   & \texttt{hotOfi1s}   & quote-update measure, 1s & \cite{cont2014price} \\
Quote & \texttt{ofi30s}  & \texttt{hotOfi30s}  & quote-update measure, 30s & \cite{cont2014price} \\
Flow  & \texttt{print}   & \texttt{hotPrint}   & latest signed print & \cite{hasbrouck1991measuring} \\
Flow  & \texttt{flow1s}  & \texttt{hotFlow1s}  & signed taker notional, 1s & \cite{biais1995empirical} \\
Flow  & \texttt{flow30s} & \texttt{hotFlow30s} & signed taker notional, 30s & \cite{biais1995empirical} \\
\midrule
\multicolumn{5}{@{}l}{\textit{Panel B: anonymous only (prices are common to all observers)}} \\
\midrule
Price & \texttt{ret1m}   & --- & one-minute midpoint return & \cite{andersen2003modeling} \\
Price & \texttt{vol1m}   & --- & 1m realized vol.\ of 1s returns & \cite{andersen2003modeling} \\
Price & \texttt{spread}  & --- & quoted spread, bp of mid & \cite{glosten1988estimating} \\
\midrule
\multicolumn{5}{@{}l}{\textit{Panel C: identity only (no anonymous counterpart)}} \\
\midrule
Flow  & --- & \texttt{hotBreadth1s}  & bucket buyer breadth, 1s\tnote{a} & \\
Flow  & --- & \texttt{hotBreadth30s} & bucket buyer breadth, 30s\tnote{a} & \\
Quote & --- & \texttt{hotShare}      & toxic share of best-quote depth\tnote{b} & \\
Quote & --- & \texttt{hotContrib}    & toxic part of best-quote imbalance\tnote{b} & \\
\bottomrule
\end{tabular}
\begin{tablenotes}[flushleft]\fontsize{7.0}{8.1}\selectfont
\item \textit{Notes:} Ten anonymous and eleven identity variables. Panel~A pairs share windows and clock; the \texttt{hot} prefix denotes cohort measures.
OFI formulas differ as described in Section~\ref{sec:prediction-features};
Panel~C has no anonymous counterpart, so no citation. Figure~\ref{fig:dropone} spells these out in words.
\item[a] Rolling sums of bucket-level distinct buyers minus sellers; wallets can recur.
\item[b] Depth denominators are clamped to at least the cohort total; see text.
\item[c] Depth within 5\,bp of the mid, in bp of implied move (Section~\ref{sec:prediction}).
\end{tablenotes}
\end{threeparttable}\end{table}

\subsection{Sampling, targets, and model fitting}
\label{sec:prediction-method}

Return forecasting needs fixed calendar horizons, whereas the venue updates on an irregular block
clock, so a block-to-block target would mix returns measured over different elapsed times. We
instead sample on a regular 100-millisecond grid, an interval close to the venue's own update
frequency because 91.3\% of adjacent blocks commit within 100 milliseconds
(Figure~\ref{fig:blocks}(b)).

Let $t$ be a grid timestamp. Features are indexed by consensus timestamps no later than $t$. Book and price
variables are states, so they take the close of the latest block committed before $t$; trade flow
and quote-update order-flow imbalance are interval measures, so at $t$ they aggregate events in the
completed interval $[t-100\text{\,ms},t)$. Recording the interval only at its endpoint prevents the
leakage that would arise if events after the forecast time entered the feature vector. For coin $c$
and horizon $\Delta$ the target is the forward midpoint return
\[
y_{c,t}(\Delta)=10^4\,\frac{m_{c,t+\Delta}-m_{c,t}}{m_{c,t}},
\]
with each midpoint the close of the latest block committed by the corresponding grid
timestamp.\footnote{The same block-close rule governs Equation~\eqref{eq:advselection}, whose
pre-event midpoint must be at most 60 seconds old and whose endpoint is missing if the state is
more than 120 seconds old, outside the sample, or inside a gate-suppressed interval. The prediction
grid carries the latest emitted state forward and does not apply these markout-specific staleness
screens. All timing is indexed by consensus timestamps on the committed tape rather than by a
separately audited local receipt-time clock.} We report horizons from 200 milliseconds to thirty
seconds and set targets that cross UTC-day boundaries to missing.

Wallet scores use July 1--10, model fitting uses July 11--20, and evaluation uses July 21--27.
Forecasting scores include orders before July 11 midnight whose ten-second markouts may
mature just after midnight; this affects the fitting boundary but precedes evaluation by ten days.
The persistence exercise instead purges the final five scoring-window minutes.
Ridge includes three unpenalized coin intercepts. Features are standardized with pooled
fitting-window means and standard deviations, including inside validation folds. Seventeen
penalties, $\lambda=10^{-7+0.5j}$ for $j=0,\ldots,16$, are compared in nine expanding folds:
each day July 12--20 is validated against all preceding fitting days. The penalty minimizing
the sum of daily validation MSEs is used to refit July 11--20.

LightGBM regression trees train on July 11--19 and validate on July 20, with no subsequent refit.
Settings are learning rate 0.05, 127 leaves, at least 5,000 observations per leaf, feature fraction
0.9, bagging fraction 0.8 each round, at most 2,000 rounds, and early stopping after 50 rounds.
The coin identifier is categorical; deterministic seeds are 20251201. At one second, early
stopping selects 192 trees for the anonymous model and 415 for the full model.

Ridge retains complete cases across the wide feature panel for all ten cohort variants and all
nine constructed targets, including unreported 60- and 120-second targets; consequently the
last 120 seconds of each day are excluded. Trees require complete cases only for the selected
feature pair and target horizon. Model pairs within each learner use common observations,
but ridge and tree samples differ slightly. Neither mask removes finite carried-forward states.
The evaluation ends July 27 following documented collector degradation and the July 28 failure;
the choice between this window and a longer evaluation was made after results were available.

We report $R^2=1-\sum_{c,t}(y_{c,t}-\widehat y_{c,t})^2/\sum_{c,t}y_{c,t}^2$ against a zero
forecast, weighting each retained grid cell equally. Inference uses daily differences
$d_j=(\mathrm{SSE}_{A,j}-\mathrm{SSE}_{A+I,j})/n_j$ and
$t=\bar d/(s_d/\sqrt{J})$, with $J=7$ July days. Day-level aggregation accommodates within-day
dependence, but precision still depends on between-day dependence. The many reported comparisons
are unadjusted, and specification selection was retrospective; the $t$ statistics are descriptive.

\subsection{Results}
\label{sec:prediction-results}

On the seven evaluation dates, the identity block alone reaches an
out-of-sample $R^2$ of 10.06\% at one second, close to the anonymous block's 10.88\%, even though
it observes only the toxic decile's activity. Combining the two raises $R^2$ to 12.31\%, a 13.2\%
relative gain with $t=9.2$. Table~\ref{tab:part2-v1} shows this is not confined to the
headline horizon: the increment remains positive through thirty seconds, beyond which
we stop reporting because the anonymous $R^2$ falls toward 2\% and relative gains on so small a
denominator are not comparable to the short-horizon cells.

\begin{table*}[tp]\centering
\begin{threeparttable}
\caption{Return prediction with the top-decile identity block (headline specification)}
\label{tab:part2-v1}
\footnotesize\setlength{\tabcolsep}{11pt}\renewcommand{\arraystretch}{1.0}
\begin{tabular*}{\textwidth}{@{\extracolsep{\fill}}lrrrrrrrr@{}}
\toprule
 & \multicolumn{4}{c}{Ridge (benchmark)} & \multicolumn{4}{c}{Gradient-boosted trees} \\
\cmidrule(lr){2-5} \cmidrule(lr){6-9}
Horizon & Anonymous & + Identity & Gain (\%) & $t$ & Anonymous & + Identity & Gain (\%) & $t$ \\
 & \multicolumn{2}{c}{$R^2$ (\%)} & & & \multicolumn{2}{c}{$R^2$ (\%)} & & \\
\midrule
0.2s & 6.91 & 8.20 & +18.7 & 8.1 & 19.26 & 21.32 & +10.7 & 5.4 \\
0.5s & 10.16 & 11.76 & +15.8 & 8.3 & 21.00 & 22.94 & +9.2 & 4.6 \\
1s & 10.88 & 12.31 & +13.2 & 9.2 & 19.48 & 20.65 & +6.0 & 5.0 \\
2s & 10.60 & 11.66 & +10.0 & 8.5 & 16.35 & 17.04 & +4.2 & 6.0 \\
5s & 9.17 & 9.79 & +6.7 & 9.5 & 12.00 & 12.36 & +3.0 & 6.1 \\
10s & 7.18 & 7.56 & +5.2 & 9.9 & 8.55 & 8.76 & +2.5 & 6.4 \\
30s & 3.56 & 3.67 & +3.0 & 6.1 & 3.74 & 3.84 & +2.7 & 1.6 \\
\bottomrule
\end{tabular*}
\begin{tablenotes}[flushleft]\footnotesize
\item \textit{Notes:} Frozen score, top decile; forward midpoint returns on the 100\,ms grid,
three model markets. Scores July 1--10, fit July 11--20, evaluation July 21--27, with trees trained
July 11--19 and July 20 held out for early stopping. Both learners use the same variable blocks.
Gain $=R^2_{\mathrm{+Identity}}/R^2_{\mathrm{Anonymous}}-1$; $t$ comes from the seven paired daily
MSE differences. Horizons past thirty seconds are omitted, the anonymous $R^2$ falling toward 2\%
there.
\end{tablenotes}
\end{threeparttable}\end{table*}

Gradient-boosted trees are the stronger learner on the anonymous features, raising their one-second
$R^2$ to 19.48\%. Identity still adds 1.17 points to reach 20.65\% ($+6.0\%$, $t=5.0$). That the
relative gain is smaller under trees than under ridge indicates that nonlinearities in the
anonymous features capture part, but not all, of what identity supplies. Under trees the gain falls
from 10.7\% at 200 milliseconds to 2.7\% at thirty seconds with daily $t$ statistics above 4 through ten seconds but only $t=1.6$ at thirty seconds.

Expanding the cohort from the top decile to the top quintile leaves the one-second ridge
relative gain nearly unchanged at 13.1\%. This supports concentrating on the upper tail,
although the primary cut was chosen retrospectively. Rolling-score results are not used as
confirmatory checks because their event-time cutoff does not separately require all markouts
to have matured before the served day.

\begin{figure}[tbp]
    \centering
    \includegraphics[width=\columnwidth]{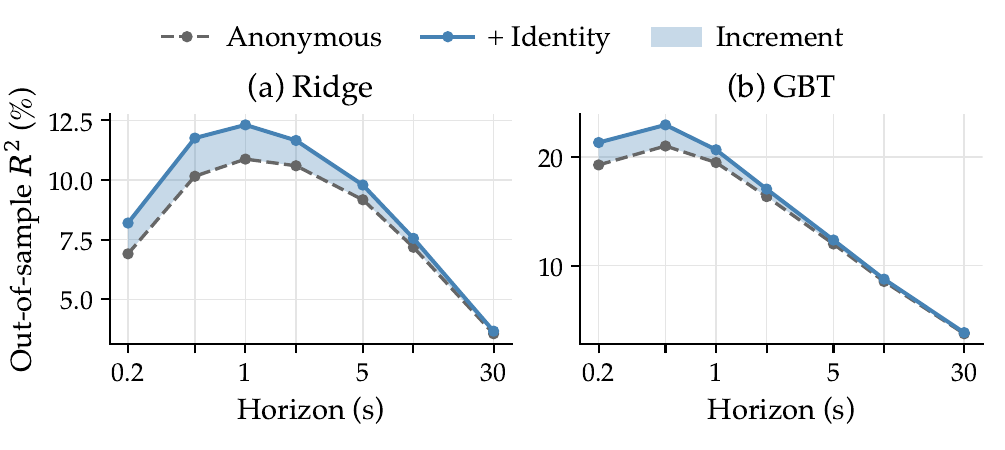}
    \caption{Out-of-sample $R^2$ by horizon, ridge and gradient-boosted trees. Shading is the
    top-decile identity increment over the anonymous benchmark.}
    \Description{Two panels of out-of-sample R-squared against horizon. Under ridge the anonymous
    benchmark peaks near 11 percent at one second and the identity-augmented model near 12.3
    percent; under gradient-boosted trees both peak near 21 to 23 percent at half a second. The
    shaded increment between them is widest at the shortest horizons and closes by thirty seconds.}
    \label{fig:gain-decay}
\end{figure}

Order-flow toxicity should matter most immediately after information enters the market and fade as
the market absorbs it, and the horizon profile in Figure~\ref{fig:gain-decay} matches that. Overall
fit is mildly hump-shaped: it rises over the first few hundred milliseconds, peaks between 0.5 and
one second, and declines thereafter. That short effective horizon is consistent with
\citet{kolm2023deep}, who find stock-specific order-book forecasts remain effective for roughly two
average price changes. The identity increment decays more steadily, its relative gain falling from
18.7\% to 3.0\% under ridge and from 10.7\% to 2.7\% under trees between 200 milliseconds and
thirty seconds. The additional predictive content is strongest at short horizons. This decay does not by
itself identify its economic source or distinguish information from anticipated order flow.

\subsection{Where the identity value lives}
\label{sec:dropone}

\begin{figure}[tbp]
    \centering
    \includegraphics[width=\columnwidth]{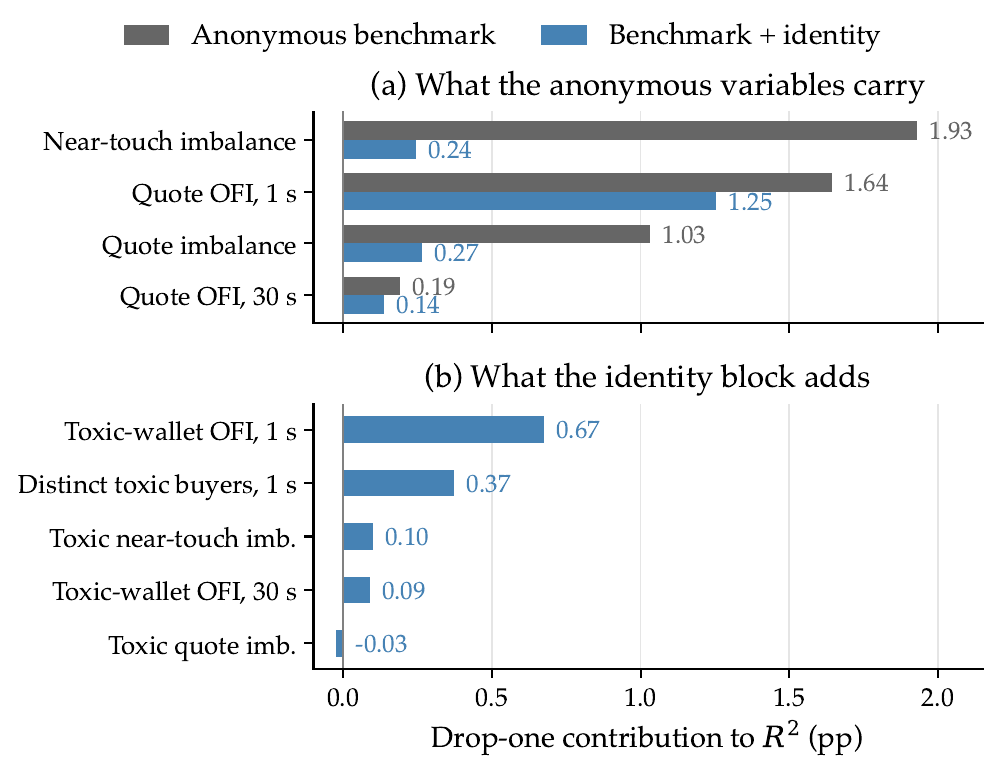}
    \caption{Drop-one contributions at the one-second ridge benchmark. (a) Anonymous predictors
    before and after adding identity; (b) identity predictors. Contributions are descriptive and
    non-additive; pairs in which neither member reaches 0.05\,pp are omitted.}
    \Description{Panel (a) shows near-touch imbalance contributing 1.93 percentage points in the
    anonymous benchmark and only 0.24 once identity is added, with quote order-flow imbalance
    falling from 1.64 to 1.25 and quote imbalance from 1.03 to 0.27. Panel (b) shows the identity
    contributions led by toxic-wallet order-flow imbalance at 0.67 and distinct toxic buyers at
    0.37.}
    \label{fig:dropone}
\end{figure}

The aggregate gain could come from an additional wallet-specific signal or from identity
variables substituting for anonymous predictors. Figure~\ref{fig:dropone} examines their overlap by
removing each variable in turn, refitting the one-second ridge model, and recording the loss in
out-of-sample $R^2$ in percentage points. Because the regressors are correlated these
contributions are descriptive and non-additive. In the anonymous benchmark the largest contribution
comes from near-touch depth imbalance (\texttt{nearImb}, 1.93); once identity is added, that falls
to 0.24. The identity-side contributions concentrate in the toxic decile's one-second quote flow
(\texttt{hotOfi1s}, 0.67) and net buyer breadth (\texttt{hotBreadth1s}, 0.37), the latter a
dimension with no anonymous counterpart at all. The pattern is consistent with overlap between
anonymous near-touch imbalance and the cohort-specific quote variables, not a causal
decomposition. Because the identity block alone reaches 10.06\% while adding only 1.43 points to
the anonymous benchmark, the two blocks largely span each other, and the natural alternative is
that identity is a lower-noise readout of the same aggregate flow rather than a distinct channel;
Section~\ref{sec:stress} weighs against the subsetting version of that alternative, since 200 cohorts
matched on size and activity, drawn from the same flow, reach at most 61\% of the increment.

\subsection{Stress tests}
\label{sec:stress}

\paragraph{Matched-wallet placebo.} Toxic wallets are unusually large and active, so identity
features built from any comparably active cohort might improve forecasts. To isolate the ranking
from size and activity we construct 200 cohorts from scored non-toxic wallets, matched to the
toxic cohort on deciles of scoring-window notional crossed with order count, rebuild all eleven
identity variables for every cohort from the block-level book and flow records, and carry them to
the same 100-millisecond grid. The exercise rebuilds each cohort over the fifty-one fitting- and
evaluation-window coin-days, or 44.0~million grid stamps and 8.8~billion cohort-observations from
126~gigabytes of replayed per-wallet quote tapes. All 201 cohorts are then scored on the same
July 21--27 evaluation pass as Table~\ref{tab:part2-v1}.

\begin{figure}[tbp]
    \centering
    \includegraphics[width=\columnwidth]{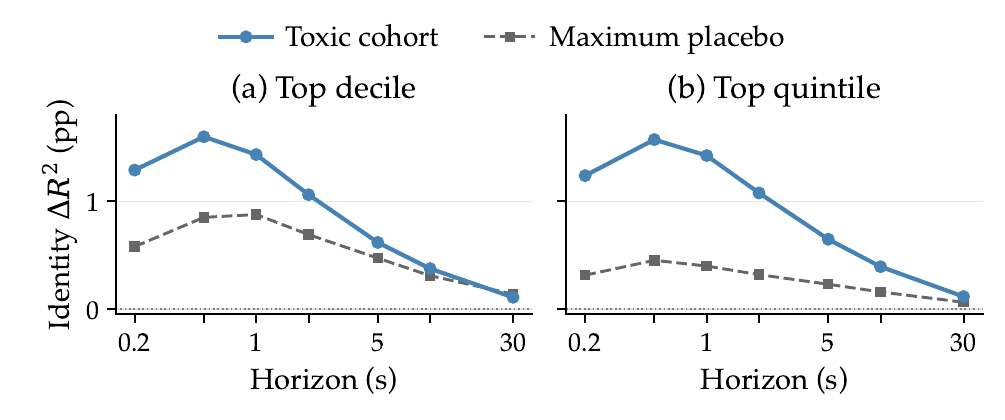}
    \caption{Identity increment for the toxic cohort against the maximum across 200
    activity-matched placebo cohorts.}
    \Description{Two panels, top decile and top quintile. In both, the toxic cohort's identity
    increment peaks near 1.6 percentage points at half a second and stays above the maximum placebo
    increment at every horizon through ten seconds, converging only at thirty seconds for the top
    decile.}
    \label{fig:placebo-hist}
\end{figure}

At one second the toxic-decile increment is 1.43 percentage points against 0.88 for the best of the
200 matched draws, so it is 1.6 times the maximum and the best draw reaches only 61\% of the real
increment. The toxic cohort exceeds every draw at every horizon through ten seconds. Figure~\ref{fig:placebo-hist} shows the separation for both tail cuts. At thirty
seconds the top decile enters the placebo range while the top quintile continues to exceed every
draw. The comparison supports a contribution beyond these activity controls, subject to the
unequal cohort sizes. Exact-cell donor limits make each draw slightly smaller than the full
toxic cohort, so the rank is descriptive and not an equal-size randomization test.

\balance
\paragraph{Feature embargo.} We test sensitivity to simulated delays by shifting \emph{every}
feature, anonymous and identity alike, back by 200 and 300 milliseconds while keeping targets
fixed. A one-second forecast then ends 1.2 or 1.3 seconds after the newest indexed information.
The relative gain falls from 13.2\% to 10.2\% to 9.1\% under ridge ($t=8.2$ under both
delays), and from 6.0\% to 4.5\% to 4.0\% under trees. With a 300\,ms delay, anonymous
$R^2$ falls from 10.88\% to 6.40\% under ridge and from 19.48\% to 9.64\% under trees.
The increment is therefore less sensitive to these delays than the anonymous benchmark. This
exercise does not audit local receipt times or establish that consensus-indexed states were
available to a trader at those times.

\paragraph{Earlier-sample comparison.} We refit the same feature specification on an independently
collected public December 2025 release \citep{albers2026openbook}: 26.3~billion messages and
21.9~million taker fills by 89,800 wallets on \textsc{btc}, \textsc{eth}, and \textsc{sol}.
Wallets are scored December 1--10, models fit December 11--20, and forecasts evaluated December
21--31. Rank persistence is $\rho=0.47$. Under the volatility-scaled specification, one-second
ridge $R^2$ rises from 12.36\% to 14.16\%, a 14.6\% relative increase ($t=3.9$).
This earlier-sample result is specification-dependent: without near-touch volatility scaling,
the one-second ridge identity increment is $-1.17$ percentage points. The current scaled December summaries
are not covered by the stored pre-scaling end-to-end audit, so a fresh panel-to-output audit
remains necessary before treating them as a fully verified replication.

\paragraph{Selection and data limitations.} The August 3 version record documents that the final
top-decile focus, near-touch volatility scaling, and July evaluation-window choice followed
available results. Chronological train/test separation prevents direct fitting on future targets
but does not remove this specification-selection effect. The prediction grid also carries emitted
states across coverage gaps and gate closures without the markout-specific staleness screens.
As a day-level sensitivity, reaggregating the fixed fitted models without July 25 and 27,
the two evaluation dates with block gaps above three, gives one-second $R^2$ of
11.72\% to 13.18\% for ridge and 19.75\% to 20.90\% for trees. This uses five days without
refitting and does not remove stale-state intervals on retained dates; a full gap-masked rerun
remains outstanding. Together with the small
number of evaluation days and unverified receipt-time availability, these limitations motivate a
new, prospectively locked evaluation with validity masks at both target endpoints.

\section{Conclusion}
\label{sec:conclusion}

Hyperliquid's wallet-attributed Level-4 record supports a direct comparison between aggregate
order-book predictors and features constructed from persistent wallet histories. Wallet markout
rankings persist across adjacent ten-day windows ($\rho=0.52$). In three active markets,
features from previously scored top-decile wallets raise one-second ridge $R^2$ from 10.88\%
to 12.31\%, and tree $R^2$ from 19.48\% to 20.65\%. Activity-matched cohort comparisons
and simulated feature delays support the increment within the tested design.

The result concerns prediction relative to a specific anonymous feature set. It does not identify
private information or establish realized trading profit. Retrospective specification choices,
carried states through missing intervals, and receipt-time availability require further validation.
A prospectively fixed test should compare richer anonymous baselines, mask invalid states, and
audit end-to-end data availability. Converting forecasting improvements into decisions additionally
requires latency, queue priority, fill probabilities, spreads, fees, and inventory constraints
\citep{avellaneda2008high}. Wallet-level cancellations and rejections offer further predictors
for that evaluation.

\section*{Ethics and Privacy}

The analysis uses public pseudonymous addresses; we do not interpret an address as a legal person,
and claims are made at the address level. Replication artifacts should minimize address-level
disclosure and avoid linkage aimed at natural persons. Identity data of this kind cut both
ways, enabling defensive screening as well as counterparty discrimination, which limits the
normative conclusions that follow from predictive accuracy alone.

\begin{acks}
Jingqiu Ding acknowledges support from the
\grantsponsor{ERC}{European Research Council}{https://erc.europa.eu/} (ERC)
under the European Union's Horizon~2020 research and innovation programme
(grant agreement No.~\grantnum{ERC}{815464}).

Daojing Zhai thanks the Yale Center for Research Computing for providing computational resources and support.
\end{acks}

\bibliographystyle{ACM-Reference-Format}
\bibliography{references}

\end{document}